\documentclass[aps,prl,groupedaddress,showpacs,preprint]{revtex4}
\usepackage{graphicx}% Include figure files
\usepackage[dvips]{epsfig}% Include figure files
\usepackage{dcolumn}% Align table columns on decimal point
\usepackage{subfigure}%
\usepackage{bm}% bold math
\usepackage{amssymb}
\usepackage{amsmath}

\begin{document}
\title{On the Geometric Principles of Surface Growth}

\author{Carlos Escudero}

\affiliation{Instituto de Matem\'{a}ticas y F\'{\i}sica Fundamental, Consejo Superior de Investigaciones Cient\'{\i}ficas, C/ Serrano 123, 28006 Madrid, Spain}

\begin{abstract}
We introduce a new equation describing epitaxial growth processes.
This equation is derived from a simple variational geometric principle and it has a straightforward interpretation in terms of continuum and microscopic physics. It is also able to reproduce the critical behavior already observed, mound formation and mass conservation, but however does not fit a divergence form as the most commonly spread models of conserved surface growth. This formulation allows us to connect the results of the dynamic renormalization group analysis with intuitive geometric principles, whose generic character may well allow them to describe surface growth and other phenomena in different areas of physics.
\end{abstract}


\maketitle

Variational principles can be found everywhere in physics.
Classical mechanics, optics, electrodynamics, and general relativity are examples of fundamental theories that are described by one such principle. These principles are not only a mathematically elegant way of describing the theory, but they are also supposed to express the fundamental physical content of it.
Quantum mechanics and quantum field theories, the most precise theories ever developed in any scientific discipline, can be expressed as quantizations (e. g. by means of path integrals) of the corresponding classical variational principles. The main role that variational principles have played in a vast range of physical theories makes them desirable when building new theoretical frameworks.

Non-equilibrium statistical mechanics is an area that has undergone an important growth in the last decades. The large number of phenomenologies found in non-equilibrium systems makes it difficult to develop unifying principles as in other fields of physics. Among all the areas of non-equilibrium statistical mechanics, few of them have the relevance and practical importance of surface growth. Together with important technological processes like thin film deposition, it describes many phenomena driven by growing interfaces in physics, biology, and other scientific fields~\cite{barabasi}. The goal of the present work is to develop a variational theory of surface growth and, in particular, in the case of epitaxial growth. Our theory will be based on a simple geometric principle and it will, at the same time, express plausible microscopic processes (which describe the adatoms dynamics) and the observed macroscopic behavior.

Epitaxial growth is characterized by the deposition of new material on existing layers of the same material under high vacuum conditions. The mathematical description uses the function $h=h(x,y,t)$, which describes the height of the growing interface in the spatial point $(x,y)$ at time $t$. Although this theoretical framework can be extended to any spatial dimension $d$, we will concentrate here on the physical situation $d=2$. A basic assumption is the no overhang approximation, which corresponds with the possibility of parameterizing the interface as a Monge patch. The macroscopic description of the growing interface is given by a stochastic partial differential equation
(SPDE) which is usually postulated using phenomenological arguments. Examples of such theories are given by the well known SPDEs named after Kardar, Parisi, and Zhang (KPZ)~\cite{kpz}, Edwards and Wilkinson (EW), Mullins and Herring (MH)~\cite{barabasi}, and Villain, Lai, and Das Sarma (VLDS)~\cite{vlds}. The last three are of the form
\begin{equation}
\label{conservation}
\partial_t h = - \nabla \cdot {\bf j} + F + \eta({\bf x},t),
\end{equation}
where the constant deposition rate $F$ will be absorbed in the definition of the function height by means of a Galilean transformation of the reference frame $h({\bf x},t) \to h({\bf x},t)+Ft$ for the reminder of this work. The spatial variable ${\bf x}=(x,y)$ and the noise term $\eta$ is a Gaussian distributed random variable that takes into account the thermal fluctuations. It has zero mean and its correlation is given by
$\left< \eta({\bf x},t)\eta({\bf x}',t') \right> = D \delta({\bf x}-{\bf x}')\delta(t-t')$,
where $D$ is the noise intensity. Eq.~(\ref{conservation}) expresses the conservation of the current ${\bf j}$, while the noise is not conserved. During years, the most general equation of this type, up to leading order, has been supposed to be
\begin{equation}
\label{complete}
\partial_t h = \nu_2 \nabla^2 h + \lambda_{22} \nabla^2 (\nabla h)^2 + \lambda_{13} \nabla (\nabla h)^3 - \nu_4 \nabla^4 h
+ \eta({\bf x},t),
\end{equation}
the discussion being restricted, mainly, to which terms need to be included, and which do not. The aforementioned SPDEs are obtained when we set to zero some of these coefficients, say, EW implies
$\lambda_{22}=\lambda_{13}=\nu_4=0$, MH corresponds to $\nu_2=\lambda_{22}=\lambda_{13}=0$, and VLDS is recovered when
$\nu_2=\lambda_{13}=0$. However, as we will show in a moment, the divergence form Eq.~(\ref{conservation}) and the leading order
Eq.~(\ref{complete}) are not the most general ways of expressing the drift term of a conserved growth mechanism.

Macroscopic equations of surface growth, as Eq.~(\ref{complete}), have been usually derived according to phenomenological arguments.
An exception to this has been the recent derivations directly from discrete microscopic models, using suitable modifications of the van Kampen system size expansion~\cite{vvedensky}. Herein we will use a different approach: we will consider the variational formulation of surface growth following the seminal idea introduced in~\cite{maritan}. In order to proceed with our derivation, we will assume that the height function obeys a gradient flow equation
\begin{equation}
\label{parisiwu}
\frac{\partial h}{\partial t}=-\frac{\delta \mathcal{V}}{\delta h}
+ \eta,
\end{equation}
where we have added the noise term, what could be thought as a Parisi-Wu stochastic quantization of the corresponding classical variational principle. The functional $\mathcal{V}$ denotes a potential that is pursued to be minimized during the temporal evolution of $h$. This potential describes the microscopic properties of the interface and of the adatom interactions and, at large enough scales, we assume that it can be expressed as a function of the surface mean curvature only:
\begin{equation}
\label{potential}
\mathcal{V}= \int f(H) \sqrt{g} d{\bf x},
\end{equation}
where $H$ denotes the mean curvature, $g$ the determinant of the surface metric tensor, and $f$ is an unknown function of $H$. We will further assume that this function can be expanded in a power series
\begin{equation}
\label{expansion}
f(H)= K+K_1 H + \frac{K_2}{2} H^2 + \cdots ,
\end{equation}
of which only the zeroth, first, and second order terms will be of relevance at large scales. The result of the minimization of the potential~(\ref{potential}) leads to the SPDE
\begin{equation}
\label{monge}
\partial_t h = K \nabla^2 h + K_1 \left[ (\partial_{xx}h)(\partial_{yy}h)-(\partial_{xy}h)^2 \right] - K_2 \nabla^4 h + \eta({\bf x},t),
\end{equation}
to leading order in the small gradient expansion, which assumes
$|\nabla h| \ll 1$~\cite{marsili}. Note that this equation possesses the same linear terms as the generic equation~(\ref{complete}), but however the nonlinearity is rather different; it is a well known nonlinear differential operator referred to as the Monge-Amp\`{e}re operator. The properties of this equation are quite remarkable. First of all, the drift is not of divergence form, but it anyway expresses mass conservation. This can be easily seen, for instance, assuming that the growth domain is the rectangle $[-L,L] \times [-L,L]$ with no flux boundary conditions. Then, integrating Eq.~(\ref{monge}) over the whole domain the linear terms vanish immediately and the nonlinear term after integrating by parts twice. The conclusion is that the integral of the function $h$ over the whole domain, this is the total mass, is conserved over time.

Apart from the conservation of mass, the scaling properties of this equation are also the desired ones. If we introduce the usual scaling $x \to e^{l}x$, $y \to e^{l}y$, $t \to e^{lz}t$, and $h \to e^{l \alpha} h$ with $l>0$, we find that scale invariance is found for the EW exponents $\alpha = 0$ and $z=2$, and so $\beta=\alpha/z=0$, because the terms proportional to $K_1$ and $K_2$ become irrelevant in the renormalization group sense. To see what are the intermediate dynamics before the EW fixed point dominates we will neglect the term proportional to $K$ and study the equation
\begin{equation}
\label{monge2}
\partial_t h = K_1 \left[ (\partial_{xx}h)(\partial_{yy}h)-(\partial_{xy}h)^2 \right] - K_2 \nabla^4 h + \eta({\bf x},t),
\end{equation}
which can be thought as the counterpart of the VLDS equation. Applying the standard dynamic renormalization group analysis~\cite{medina} to this SPDE we arrive at the following renormalization group flow equations at one loop order
\begin{eqnarray}
\label{k2}
\frac{d K_2}{dl} &=& K_2 \left( z-4 -\frac{K_1^2 D}{16 \pi K_2^3}\cos^2(\theta) [2+\cos(2\theta)] \right), \\
\frac{d K_1}{dl} &=& K_1 (\alpha-4+z), \\
\frac{d D}{dl} &=& D (z-2-2\alpha),
\end{eqnarray}
where $\theta$ is the angle formed by the wavevector and the abscise axis, which parameterizes the surface in radial coordinates. From here we can straightforwardly read that scale invariance is reached for the values of the critical exponents $\alpha=2/3$, $z=10/3$, and $\beta=1/5$, this is,
Eq.~(\ref{monge2}) is in the VLDS universality class. The renormalization group analysis clearly shows that the term proportional to $K_2$ is irrelevant (up to one loop order), and that the interface dynamics is dominated by the $K_1$ term in the long time limit, a result identical to the one obtained for the
VLDS equation~\cite{vlds}. There is, however, one important difference among both equations respecting to the renormalization of $K_2$. For the VLDS equation, the renormalization group flow~(\ref{k2}) approaches a positive fixed point, a situation reminiscent to that of the $1D$ KPZ equation~\cite{kpz}. In the KPZ situation, however, the interface dynamics is a nontrivial combination of diffusion and nonlinearity, while in the VLDS one the nonlinearity dominates. In the present case, except for $\theta=\pi/2,3\pi/2$, Eq.~(\ref{k2}) approaches zero with an increasing velocity till eventually blowing up $dK_2/dl \to -\infty$ when $K_2 \to 0$. So this point is the end of the classical evolution of the differential equation and, in our current view, the end of the physical evolution too. Anyway, the point $K_2=0$ is a branching point, and one could consider extensions of the solution to a different branch, negative or complex (allowing negative branches will rend the theory unstable and one would have to continue expansion~(\ref{expansion}) in order to recover stability~\cite{vvedensky2}). But we are not going to consider such extensions as physical within this context. If $\theta =\pi/2,3\pi/2$, then $K_2(l)$ simply falls exponentially to zero. These results explain how the nonlinearity becomes dominant and the diffusion irrelevant in the hydrodynamic limit. One may wonder about the $\theta$ dependence in Eq.~(\ref{k2}). It means that the renormalization of diffusion constant $K_2$ depends on the nonlinearity, which is actually the small gradient approximation of the surface Gaussian curvature (see below). Diffusion is only affected by curvature along the direction of diffusion, and thus the explicit dependence on $\theta$ after fixing the coordinate system. Of course, the result becomes independent of this angle asymptotically in the scale $l$, as we have already seen.

So we see that if Eq.~(\ref{monge}) describes an interface that initially fluctuates following MH dynamics (this is, $\alpha=1$,
$z=4$, and $\beta=1/4$), then it will switch to the VLDS universality class for intermediate times, and finally, in the long time, will behave as the EW equation. This same transient of scaling behavior was found in~\cite{vvedensky2} by means of a formidable renormalization group calculation of the $2D$ activated surface diffusion model, except for the final EW behavior. These authors also find the MH initial condition for the renormalization group flow equations, preceded in certain temperature ranges by a short transient characterized by conserved MH dynamics. Of course, this initial transient cannot be found in our theory, since we are not considering conserved fluctuations. One obvious improvement would be to include both conserved and nonconserved noise in a generalization of the Parisi-Wu quantization~(\ref{parisiwu}). The asymptotic EW dynamics was found in the $2D$ EW discrete model, in which a special type of adatom diffusion is considered~\cite{vvedensky2}. Presumably, combining both models a similar output to ours could be found.
Herein we have seen that such evolutions have a clear geometrical meaning: short times are characterized by the minimization of the square of the mean curvature, during intermediate times the mean curvature itself is minimized, while in the long time limit the zeroth power of the curvature, which corresponds to the surface area, is minimized. These dynamical behaviors are in correspondence with the MH, VLDS, and EW universality classes respectively. Furthermore, this theoretical framework allows us to continue the series expansion in terms of powers of the curvature to include cubic or higher orders, and so to describe shorter spatiotemporal scales, at least in principle.
Practically it might difficult to perform this program up to an arbitrary order in the expansion due to the possible presence of nonuniversal features when the spatial scales become of the same order of magnitude of the lattice spacing.

Another characteristic feature that has been observed in both numerical simulations and experiments of epitaxial growth is mound formation~\cite{vvedensky}. Eq.~(\ref{monge}) is able to reproduce this feature too. The terms proportional to $K$ and $K_2$ are stabilizing, and push the surface towards planar profiles. On the other hand, the nonlinearity proportional to $K_1$ favors mound formation. The Gaussian curvature of the surface is given by
\begin{equation}
\mathcal{K}_G =
\frac{(\partial_{xx}h)(\partial_{yy}h)-(\partial_{xy}h)^2}{1 +
(\nabla h)^2},
\end{equation}
which is of the same sign as the $K_1$ term. This shows that those parts of the interface provided with a positive Gaussian curvature grow, while those provided with a negative Gaussian curvature decrease in time. This indicates how mounds alone might form, contrary to linear instabilities that favor both negative and positive curvatures, and thus mounds and valleys formation. To further clarify how mounds appear, we will consider the evolution equation for just the nonlinear term
\begin{equation}
\partial_t h = K_1 \left[ (\partial_{xx}h) (\partial_{yy}h) - (\partial_{xy}h)^2
\right].
\end{equation}
By means of a variables separation $h(x,y,t)=a(t)b(x,y)$ we get
\begin{equation} \label{separation} \frac{a'}{K_1 a^2} =
\frac{(\partial_{xx}b) (\partial_{yy}b) -
(\partial_{xy}b)^2}{b}=C,
\end{equation}
for some constant $C$. The solution for the temporal variable reads
\begin{equation}
a(t) = \left\{ \left[ a(0) \right]^{-1} - C K_1 t \right\}^{-1},
\end{equation}
showing that for $C<0$ it decays in time, while for $C>0$ it grows unboundedly till it blows up in finite time. Of course, in the real situation blow up will not occur due to the underlying lattice structure. Because mass is conserved, the blow up appears as the consequence of the collapse of a microscopic structure. Such a collapse is precluded by the existence of a finite lattice spacing, and it is affected by the regularizing effects of the $K$ and $K_2$ terms. By Eq.~(\ref{separation}) we find, as expected, that the sign of the constant $C$ is the same that the sign of the Gaussian curvature of the surface. We only need to show that there exist nontrivial positive solutions to the equation
$(\partial_{xx}b) (\partial_{yy}b) - (\partial_{xy}b)^2 = C b$, for some positive constant $C$. The existence of such a solution was rigorously proven in~\cite{charro}, what opens the possibility of this mechanism of mound formation. We do not know if there exist nontrivial solutions to this equation for some negative values of the constant $C$, but this fact is physically unimportant, as these solutions, provided they exist, will approach the trivial one monotonically in time. One important point here is that the nonlinear Monge-Amp\`{e}re term is able to promote both mound formation and VLDS critical dynamics. Both features were found simultaneously in heteroepitaxial growth of InP layers~\cite{sarma}, and so this mechanism constitutes a plausible explanation of the observed phenomenology. Note that we can consider Eq.~(\ref{monge2}) as a possible continuum description of the $2D$ activated surface diffusion model (which describes homoepitaxial growth) or the heteroepitaxial growth experiments~\cite{sarma}. On the other hand, the full equation~(\ref{monge}) can be considered as the description of the epitaxial system after a suitable control protocol, incorporating EW diffusion, has been introduced. The resulting dynamics will approximate asymptotically the EW universality class, and thus the desired the smooth (or more precisely logarithmically rough) interface.

The variational formulation allows us to derive the stationary probability functional of the interface configuration. In the small gradient approximation the potential functional reads
\begin{equation}
\mathcal{V}[h({\bf x},t)]= \int \left[ -\frac{K}{2} (\nabla h)^2 + K_1 h_x h_y h_{xy} - \frac{K_2}{2} (\nabla^2 h)^2 \right] dx dy,
\end{equation}
and so, the probability of a given spatial configuration $h({\bf x})$ in the long time limit is given by
\begin{equation}
\mathcal{P}_s [h({\bf x})]= \mathcal{N} \exp \left \{- 2D^{-1}
\mathcal{V}[h({\bf x})] \right\},
\end{equation}
where the normalization constant is
\begin{equation}
\mathcal{N}^{-1}= \int \mathcal{D}h({\bf x}) \exp \left \{- 2D^{-1}
\mathcal{V}[h({\bf x})] \right\},
\end{equation}
and the functional integral extends to all the functions $h({\bf x})$ satisfying the prescribed boundary conditions. Another advantage of Eq.~(\ref{monge}) is its easy interpretation in terms of microscopic physical processes, this is, the underlying adatoms dynamics. The term proportional to $K$ is simply diffusion, and correspondingly denotes the random walk character of the adatom motion on the surface. The term proportional to $K_2$ is microscopically related to fine properties of the random walk. Such a term may arise, for instance, when studying corrections to the central limit theorem due to a small scale underlying structure. It is clear now why it is the first in becoming irrelevant when subject to renormalization group flow. The nonlinear term, proportional to
$K_1$, expresses optimal mass rearrangement: the new deposited adatoms, which form a disordered structure, move towards the new, more favorable flat disposition that minimizes the interface chemical potential. This rearrangement is such that the square of the distance among the initial and final adatoms position is minimized~\cite{evans}. The reason why the squared distance, and not the distance itself, is minimized has to do with the random dispersal properties of the adatoms. As they move randomly, they propagate along areas rather than distances. Note that this diffusive process could be identified with the driving force generating mounds in the experimental situation described in~\cite{sarma}. It is remarkable that a similar process of optimal mass rearrangement has been found in atmospheric physics, particularly in semigeostrophic flow~\cite{cnp}.

The nonlinear term in Eq.~(\ref{monge}) was derived from a specific variational principle, but it appears in much more general situations. It can actually be derived from an infinite number of inequivalent potential functionals, a fact specially relevant in the context of high energy physics~\cite{fairlie}. Apart from its interesting geometrical properties (it is related to both mean and
Gaussian curvatures of the surface, as we have seen), it also appears in the theory of optimal transport~\cite{evans} and in this context is related to atmospheric air flow~\cite{cnp}. Another interesting property of Eq.~(\ref{monge}) is its connection with the $1D$ KPZ equation: assuming the solution form $h(x,y,t)=-y u(x,t)$, we see that it obeys the $1D$
KPZ equation after including $y$ dependence in the coefficient of the nonlinear term and setting $K_2=0$. Despite the purely mathematical nature of this symmetry, the generic character of both equations may reveal its physical significance in a field different from surface growth.

In conclusion, we have derived a stochastic growth equation from a simple variational principle. This equation describes the physics of the growing interface at both the macro and micro levels, and reproduces mound formation and the known critical exponents. Such a complete and unified approach has not been achieved previously, to our knowledge, in the epitaxial growth literature. This is so possibly due to the omnipresent assumption of the divergence form
Eq.~(\ref{conservation}) for models describing epitaxial growth.
In fact, this assumption has been adopted for years as the only way of assuring the conservation of mass; however, as we have shown, mass is conserved in models not belonging to this class.
Furthermore, Eq.~(\ref{monge}) can be derived from simple and general principles, what makes it a very generic construction of potential influence in a wide range of different areas within physics. As we have already mentioned, high energy, atmospheric, condensed matter, and statistical physics can be connected to it. Indeed, we are persuaded that the profoundly generic nature of this construction will make it appear in many other different areas as well.

This work has been supported by the MEC (Spain) through Project No. FIS2005-01729.

\end{document}